%% file: main.tex
\documentclass[letterpaper,twocolumn,10pt]{article}
\usepackage{usenix-2020-09}

\input{preamble}

\begin{document}

\date{}

\title{\Large \bf \thiswork{}: Towards an Open and Compiler-centric Ecosystem \\for GPU Kernel Performance Tooling on Modern AI Workloads}

\author{
{\rm Yue Guan$^{1\dagger}$, Yuanwei Fang$^{2\ddagger}$, Keren Zhou$^{3,4}$, Corbin Robeck$^{2\ddagger}$,}\\{\rm Manman Ren$^{2\ddagger}$, Zhongkai Yu$^{1\dagger}$, Yufei Ding$^{1,2\dagger}$, Adnan Aziz$^{2\ddagger}$}\\
$^1$University of California, San Diego, $^2$Meta, $^3$George Mason University, $^4$OpenAI\\
$^\dagger$\{yueguan, zhy055, yufeiding\}@ucsd.edu\\
$^\ddagger$\{fywkevin, robeck, mren, adnanaziz\}@meta.com\\
kzhou6@gmu.edu
}

\maketitle


\input{section/0_abstract}
\input{section/1_introduction}
\input{section/2_motivation}
\input{section/3_compiler_centri_design}

\input{section/4_profiler_tool}

\input{section/5_evaluation}
\input{section/6_discussion}

\bibliographystyle{plain}
\bibliography{reference}

\appendix

\input{section/artifact.tex}

\end{document}

%% file: preamble.tex
\newcommand{\R}[1]{{#1}}
\newcommand{\OURR}[1]{{#1}}

\newcommand{\thiswork}{\textsf{KPerfIR}}

\newcommand{\code}[1]{\texttt{#1}}

\newcommand{\Sec}[1]{Sec.~{#1}}
\newcommand{\Fig}[1]{Fig.~{#1}}
\newcommand{\Tbl}[1]{Tbl.~{#1}}

\usepackage{graphicx} 
\usepackage{amsmath}

\usepackage{lipsum}

\usepackage[strict]{changepage}
\usepackage{framed}
\definecolor{quoteshade}{RGB}{220,220,220}
\definecolor{quoteframe}{RGB}{160,160,160}

\usepackage{makecell}

\usepackage{booktabs}

\usepackage{array}

\usepackage{makecell}

\usepackage{circledtext}
\newcommand{\ballnumber}[1]{\circledtext*[charf=\Large]{#1}}

\renewcommand{\paragraph}[1]{\vspace{0.1cm}\noindent\textbf{#1}}

\usepackage{capt-of}

\usepackage{multirow}

\usepackage{wasysym}

%% file: section/0_abstract.tex
\begin{abstract}
In this work, we propose \thiswork{}, a novel multilevel compiler-centric infrastructure to enable the development of customizable, extendable, and portable profiling tools tailored for modern artificial intelligence (AI) workloads on modern GPUs. 
Our approach integrates profiling capabilities directly into the compiler workflow, allowing profiling functionalities to be implemented as compiler passes, offering a programmable and reusable framework for performance analysis. 
This design bridges the gap between compilers and profilers, enabling fine-grained insights into complex optimization challenges such as overlapping the execution of fine-grained function units on GPUs.
\thiswork{} is integrated into the Triton infrastructure to highlight the power of a compiler-centric approach to advance performance analysis and optimization in the ever-evolving landscape of AI compilers.
Our evaluation shows that our tool incurs low overhead (8.2\%), provides accurate measurements (2\% relative error), and delivers actionable insights into complicated GPU intra-kernel optimizations.
\end{abstract}

%% file: section/1_introduction.tex
\section{Introduction}
In the era of artificial intelligence (AI)\cite{lecun2015deep}, the popularity of AI compilers has increased dramatically~\cite{chen2018tvm, ansel2024pytorch}.
Modern AI compilers, such as Triton\cite{tillet2019triton}, have become popular in bridging the gap between high-level machine learning framework operators (e.g. general matrix multiplications (GEMMs)\cite{nvidia2023cublas}, softmax\cite{rumelhart1986learning}, etc.) and the low-level, target-specific machine code\cite{NVIDIAPTX}. 
The diversity of workloads and the rapid evolution of hardware architectures, particularly in GPUs, pose significant challenges in developing frameworks that are both modular in development and performant ``out of the box'' for users.
Addressing these demands requires a flexible compiler infrastructure, like LLVM\cite{lattner2004llvm}'s Multi-Level Intermediate Representation (MLIR)\cite{lattner2021mlir} with a modular design, and Triton for its customizable high-performance AI operators for 
diverse workloads and backends.

\begin{figure*}
    \centering
    \includegraphics[width=\linewidth]{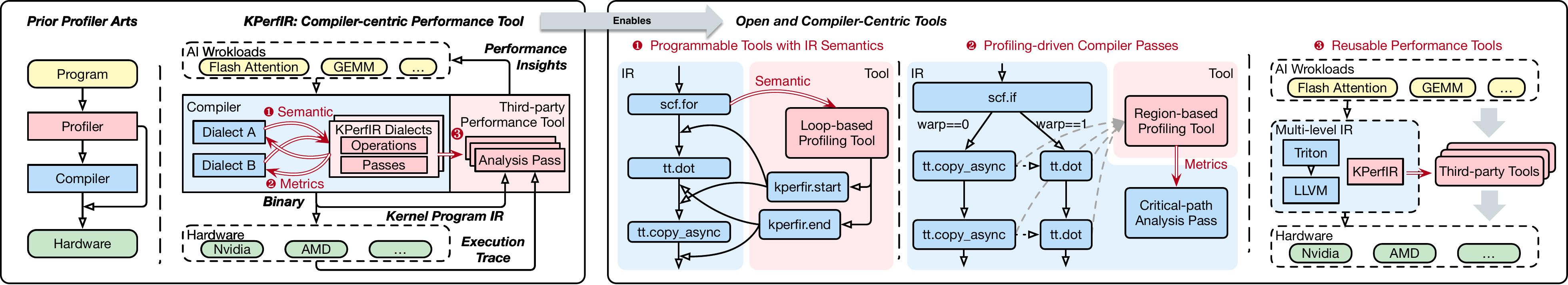}
    \vspace{-0.7cm}
    \caption{\R{Concept of the \thiswork{} infrastructure and ecosystem for compiler-centric performance tool. (Left) Overview and comparison of \thiswork{}'s compiler-centric design and prior profiler designs. (Right) Demonstrative examples of novel performance tools facilitated by the compiler-centric design of \thiswork{}.}}
    \vspace{-0.3cm}
    \label{fig:concept}
\end{figure*}

Despite significant advancements, existing compilers often struggle to outperform hand-tuned implementations like cuBLAS\cite{nvidia2023cublas}, rocBLAS\cite{amd2023rocblas}, and CUTLASS\cite{Thakkar_CUTLASS_2023}.
This limitation stems from GPU architectures' rapid evolution and the continuous emergence of novel operator variants. 
For instance, Nvidia’s Hopper architecture incorporates advanced acceleration units such as $5^{\text{th}}$ generation Tensor Cores (TC)\cite{nvidia2018turing} and Tensor Memory Accelerators (TMA)\cite{choquette2023nvidia}, which require sophisticated compiler passes to fully leverage their potential with a dataflow-oriented programming paradigm\cite{choquette2023nvidia}. 
Meanwhile, the latest Flash-Attention-3\cite{shahflashattention} (FA3) kernel, crucial for large language models (LLMs)\cite{brown2020language, llm_for_finance, llm_for_medical}, employs complex tiling and pipelining techniques that traditional compilers cannot effectively handle.

To enhance the compilers with novel compute paradigms and hardware features, profiling tools are critical in developing high-performance AI kernels and optimization passes. 
They are essential for identifying performance bottlenecks, analyzing execution flow, and understanding memory access patterns.
Developers depend on these tools to pinpoint instruction stalls, optimize kernel performance, and refine compiler passes. 
Even for feedback-guided auto-optimization compilers\cite{zheng2020ansor}, the design space consists of manually articulated transformation primitives and requires precise performance insights to guide effective optimization. 
Without effective profiling tools, achieving peak performance on modern GPU platforms becomes a daunting challenge.

However, existing profilers are poorly aligned with the advancements of compiler infrastructure to assist in the development of themselves and AI kernels.
\textbf{Our key observation is that prior profiler designs are isolated from the compiler system and lack key connections to the upper-level framework operations.}
Profiling tools are usually developed as external tools, detached from the intricacies of the compilation process, and fail to provide framework operation-informed, actionable insights tailored to the unique challenges of AI compilers. 
This disconnection severely hinders the ability of operator and AI compiler developers to enhance the performance of ML workloads.

To address this, we propose a novel compiler-centric infrastructure, \thiswork{}, to facilitate the development of customizable and reusable performance tools\footnote{\OURR{The source code is open sourced at \url{https://github.com/triton-lang/triton/tree/main/third_party/proton/dialect}}.}, as illustrated in \Fig{\ref{fig:concept}}.
Our approach establishes a seamless interplay between the compiler and profiler, enabling the exchange of program semantics and profiling metrics. 
\OURR{
Central to our methodology is incorporating multi-level compiler IR instrumentation into the profiling workflow. 
This allows performance tools to be composed as compiler passes with compiler-integrated profiling operations.
This design fundamentally enhances performance tooling in three key ways.
}

\textbf{\ballnumber{1} Programmable tools with IR semantics.} Existing profilers have limited information on the program, restricting their ability to capture performance details tied to program semantics. For example, traditional tools cannot track how software pipelining\cite{huang_alcop_2023} overlaps and evolves across pipeline stages due to missing loop-level information. By integrating IR-level instrumentation, \thiswork{} enables profiling tools to capture these nuanced behaviors effectively.

\textbf{\ballnumber{2} Profile-driven design of compiler passes.} Many optimization techniques, such as autotuning\cite{tunner1, tunner-autotvm}, rely on performance feedback to guide their tuning processes\cite{zheng2020ansor}. Traditional approaches treat compilers and profilers as isolated sub-modules, requiring standalone systems to coordinate optimization and evaluation. With \thiswork{}, profiling passes can directly interact with compiler optimization passes, streamlining the feedback loop and enabling more optimizations.

\textbf{\ballnumber{3} Reusable and extendable performance tools.} By implementing tools native to the compiler IR, \thiswork{} ensures their reusability and portability across frameworks and backends. Profiling tools can operate on shared upper-level representations used by AI frameworks and seamlessly lower to diverse backends, including Nvidia and AMD GPU platforms. Building on the MLIR philosophy of a shared set of infrastructure, we provide not only a powerful set of analysis tools but also a common set of building blocks and infrastructure for a community-driven ecosystem.

To showcase the capabilities of the \thiswork{} infrastructure, we present the first region-based timing tool for GPUs to provide insights into intra-kernel behaviors. 
By leveraging the program's IR semantics, the tool efficiently updates the profiled regional timestamps to recover the inaccuracy caused by instrumentation, ensuring accurate profiling results while minimizing memory overhead. 
This level of precision was previously difficult to achieve because traditional profilers lack access to compiler-generated semantic information, such as loop structures and regional boundaries, making it challenging to attribute execution metrics accurately to high-level behaviors. 
This timing tool plays a crucial role in optimizing modern GPU kernels by enabling precise analysis of fine-grained overlapping and presenting an intuitive timeline visualization to guide performance optimizations.

We conducted an in-depth case study on the FA3 kernel. 
Leveraging the tool's fine-grained profiling results, we identified idle bubble regions in the baseline implementation and extracted key optimization insights. 
Based on these findings, we implemented novel compiler passes to enhance the kernel with an improved overlapping strategy, effectively reducing performance bottlenecks.
In addition, we introduced performance modeling based on the profiling results to evaluate the efficiency of the optimized overlapping design.

This work makes the following key contributions:
\begin{itemize}
    \itemsep0em 
    \item We propose \thiswork{}, a compiler-centric infrastructure enabling customizable performance tools. Integrated into the Triton compiler, it supports AMD and Nvidia platforms, providing essential support for developers optimizing AI compilers and operators (\Sec{\ref{sec:compiler_centric}}).
    \item As a demonstration of the proposed infrastructure, we develop a novel region-based timing tool. This tool exemplifies the compiler-centric approach and offers fine-grained GPU profiling capabilities (\Sec{\ref{sec:tool}}).
    \item We perform an in-depth case study using the proposed timing tool to analyze GPU overlapping. The study reveals valuable insights into intra-kernel overlapping techniques, such as warp specialization (\Sec{\ref{sec:evaluation}}).
\end{itemize}

%% file: section/2_motivation.tex
\section{Background and Related Works} \label{sec:background}
Although proposed methodologies support both Nvidia and AMD GPUs, we use the ones from Nvidia without any loss of generality unless discussing AMD platform-specific features.
In the following, we describe the background of GPU compilers and profilers, focusing on their use in AI workloads.

\subsection{GPU Compilers}

The design of AI compilers has progressed significantly, with the introduction of MLIR\cite{lattner2021mlir} marking a pivotal step toward modularity, extensibility, and multi-layered abstraction. 
MLIR simplifies the complexity of modern computational workloads with a reusable framework that lowers the learning curve for compiler development while enabling custom components and third-party tool integration to drive innovation.

Building on the foundation of MLIR, Triton\cite{tillet2019triton} is a groundbreaking compiler designed to bridge AI workloads and GPU hardware. 
With dialects specifically tailored for GPU programming, Triton offers high-level abstractions that simplify the development of efficient GPU kernels, addressing the needs of computationally intensive AI applications. 
Triton leverages MLIR’s modular architecture to achieve reusability, seamless integration, and multi-level optimizations, exemplifying how domain-specific needs can be addressed within a unified framework.
Triton's Intermediate Representations (IRs) encompass various dialects, including TTIR, TTGIR, Triton, TritonGPU, TritonNvidiaGPU, and TritonAMDGPU, each catering to specific stages of GPU programming.

\begin{table}[]
    \caption{\R{Comparison of GPU performance tools}}
    \vspace{0.3em}
    \resizebox{\linewidth}{!}{
    \begin{tabular}{lcccc}
    \toprule
    Tools                          & Program IR & \makecell{Intra-Kernel\\Profiling} & \makecell{Customized\\Regions} & \makecell{Platform\\Portability} \\ \hline
    \rowcolor[HTML]{EFEFEF} 
    NCU\cite{NVIDIANsightCompute}                            &  \Circle          &            \Circle            &      \Circle              &        \Circle              \\ 
    RocTracer\cite{ROCmROCTracer}                      &   \Circle         &   \Circle                     &     \Circle               &     \Circle                 \\ 
    \rowcolor[HTML]{EFEFEF} 
    AMD ATT                        &   \Circle         &      \CIRCLE                  &      \Circle              &    \Circle                  \\ 
    TorchProfiler\cite{torchprofiler}                  &    \Circle        &   \Circle                     &     \Circle               &    \CIRCLE                  \\ 
    \rowcolor[HTML]{EFEFEF} 
    Mosaic Profiler\cite{jax2018github}                &    \Circle        & \CIRCLE                       &    \CIRCLE                &    \Circle                  \\ 
    TK Profiler\cite{spector2024thunderkittens}                    &  \Circle          &    \CIRCLE                    &     \CIRCLE               &    \Circle                  \\ 
    \rowcolor[HTML]{EFEFEF} 
    Tool with \thiswork{} (\S\ref{sec:tool}) &    \CIRCLE        &         \CIRCLE               &      \CIRCLE              &   \CIRCLE                   \\ \bottomrule
    \end{tabular}
    }
    \vspace{-1em}
    \end{table}

\subsection{GPU Profilers}

GPU performance tools are indispensable for analyzing workload execution characteristics and identifying performance bottlenecks. 
They enable developers to understand application behavior and hardware utilization, which are critical for achieving optimal performance on GPU platforms. 
The GPU performance tools can be divided into two major categories.

\R{
\paragraph{Performance tools} are ready-to-use profilers include NVIDIA’s Nsight Compute (NCU)\cite{NVIDIANsightCompute} and Nsight Systems (NSys)\cite{NVIDIANsightSystems}, as well as AMD’s RocTracer and RocProfiler\cite{ROCmROCProfiler}. NCU provides detailed kernel-level analysis for CUDA applications, offering various metrics and actionable optimization suggestions. NSys delivers a holistic view of system-wide performance, capturing interactions between CPU and GPU to diagnose latency and synchronization issues. Similarly, RocTracer\cite{ROCmROCTracer} supports performance optimization for AMD GPUs by tracing and analyzing applications in heterogeneous environments. An important aspect of these profiling tools is timing start and stop calls (e.g., cudaProfilerStart and cudaProfilerStop), which are implemented as host-side APIs \textit{around} kernel launch calls and not directly within kernel code. 
}

\R{
MosaicGPU profiler\cite{jax2018github}, which inserts PTX instructions using high-level Python bindings, represents the closest idea to our approach. However, it operates at the assembly code level rather than on the high-level IRs, restricting its ability to provide comprehensive and reusable profiling capabilities.
ThunderKitten (TK)\cite{spector2024thunderkittens}, a promising DSL for the Nvidia platform, also developed a custom timing interface recently for region-based tracing.
However, this tool is coupled with its DSL design and only supports the Nvidia platform.
}

\paragraph{Profiling infrastructures} like CUPTI\cite{nvidia2023cupti}, NVBit\cite{villa2019nvbit}, RocProfiler offer some level of instrumentation for customization. 
They allow developers to insert custom instructions at runtime, providing flexible profiling options. 
However, they are tightly coupled to the program and instrumentation framework\cite{guan2024amanda}.
This approach lacks portability and fails to integrate seamlessly with compiler workflows, limiting their ability to support cross-platform development.

\subsection{GPU Overlapping Techniques}

To effectively utilize GPU devices' resources and unleash their full potential, GPU kernels have to arrange diverse execution and memory units to overlap their execution to achieve better concurrency.
With the introduction of dedicated on-chip acceleration units, such as Tensor Core\cite{choquette2018volta} and Tensor Memory Accelerator\cite{choquette2023nvidia}, their overlapping is getting much more significant and difficult.

\begin{figure}
    \centering
    \includegraphics[width=\linewidth]{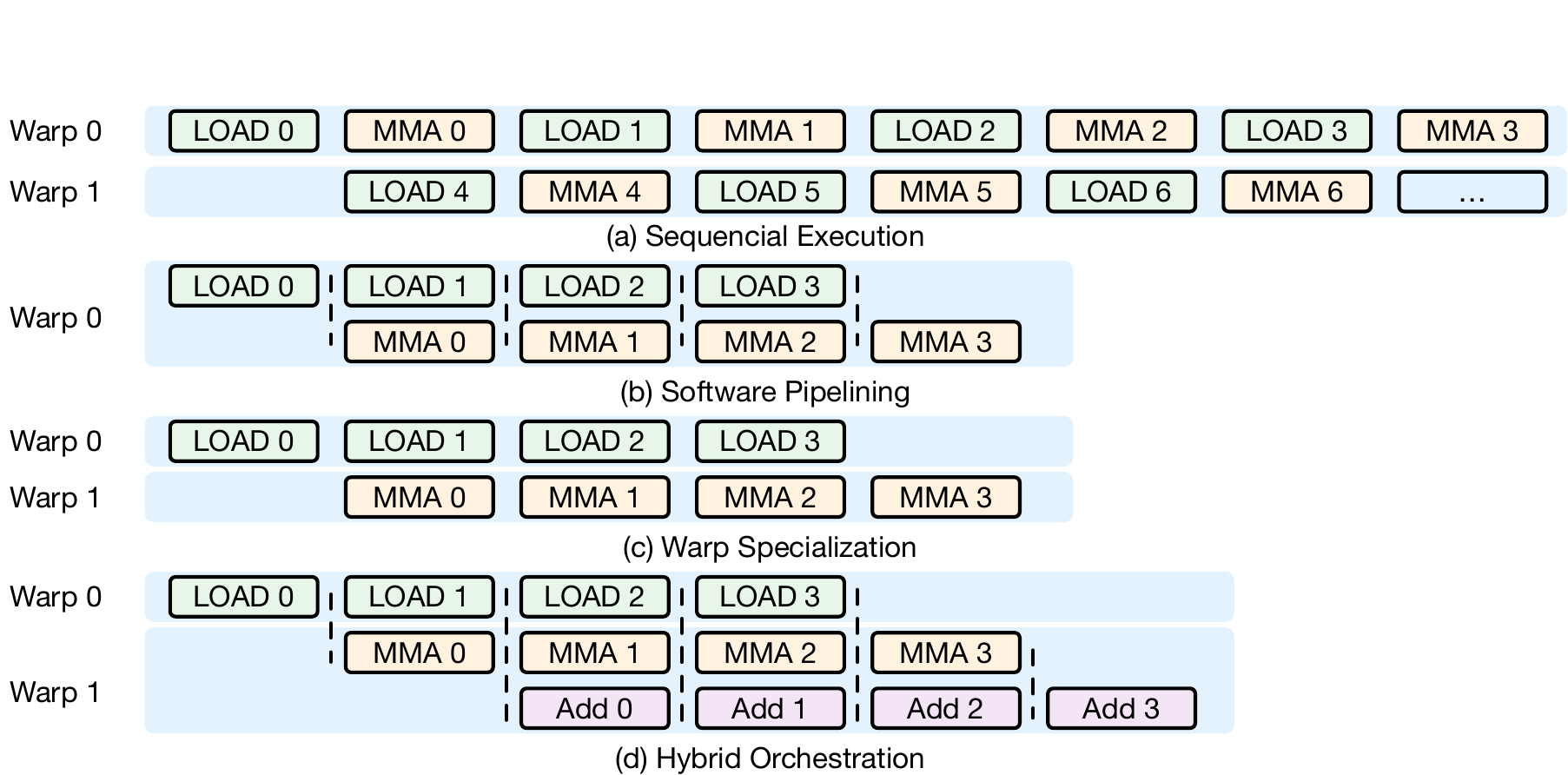}
    \vspace{-0.8cm}
    \caption{GPU overlapping techniques}
    \vspace{-1em}
    \label{fig:orchastration}
\end{figure}

\paragraph{Wave Priority and Instruction Scheduling.} 
The most straightforward solution for overlapping is to extend the scheduler with explicit priority setting.
This is implemented as the wave priority technique mainly adopted by the AMD platform through compiler-based intrinsics (e.g., LLVM's \code{\_\_builtin\_amdgcn\_s\_setprio()}), which controls how wavefronts are scheduled and executed on the Compute Units (CUs) at compile-time. The CU will, at runtime, choose the wavefront with the higher priority when scheduling conflicting instruction types. By setting well-suited wave priorities, programmers can fully utilize the execution pipeline by explicit overlapping of memory loads on one wave and compute operations on another \cite{ck}. Overlapping can be further refined through compiler scheduling intrinsics such as \code{\_\_builtin\_amdgcn\_sched\_barrier()} to control the clustering and pipelining of matrix operations.

\paragraph{Software Pipelining (SWP)} transforms the execution of independent loop iteration operations (i.e., memory and compute) into multiple stages to overlap between iterations, as shown in \Fig{\ref{fig:orchastration}}-(b). 
By using asynchronous instructions for data loading~\cite{bauer_cudadma_2011} and computation, different operations can be executed concurrently to overlap the latency. 
However, SWP requires extra scratchpad storage and registers in the loaded memory hierarchy to store the intermediate results for each stage.
Software pipelining is usually adopted at multiple levels, such as global memory to shared memory and shared memory to register files, to hide the latency effectively.
Besides the use for overlapping data loading and computation, SWP is also used for overlapping different sections of computations executed by separate hardware units.
VALU, VMEM, and SMEM instructions\cite{ck, amd_isa} can all be scheduled independently - the goal is to have these run entirely in parallel to tensor/matrix core operations.
Due to its key advantage of enhancing utilization, SWP has gained widespread acceptance in high-performance operator libraries\cite{Thakkar_CUTLASS_2023} and compiler optimizations\cite{huang_alcop_2023}.

\paragraph{Warp Specialization (WS)} is another overlapping technique built with the producer-consumer model\cite{bauer_cudadma_2011,MLSYS2023_d6cde2c1}.
After splitting execution into stages, WS assigns stages to dedicated warps as producers and consumers.
Instead of overlapping between different stages of loop iterations, WS overlaps between producer and consumer stages in different warps as depicted in \Fig{\ref{fig:orchastration}}-(c).
WS was previously proposed as a solution for irregular and large problems to overlap unbalanced workloads and reduce register pressure\cite{crago_wasp_2024}.
Compared to SWP, where each warp acts as a producer and consumer, WS separates the roles so that registers can be assigned properly.
For example, the data loading stage consumes much fewer registers compared to the heavy-lifting computation stage.
However, this feature was not fully utilized due to the lack of hardware warp-level register allocation support.
Recently, on the Nvidia Hopper architecture\cite{choquette2023nvidia}, novel features like register reallocation and asynchronous transaction barriers make WS achieve superior overlapping performance\cite{shah2024flashattention, Thakkar_CUTLASS_2023}.

\vspace{-0.5cm}
\section{Motivation} \label{sec:motivation}

\begin{figure}
    \centering
    \includegraphics[width=\linewidth]{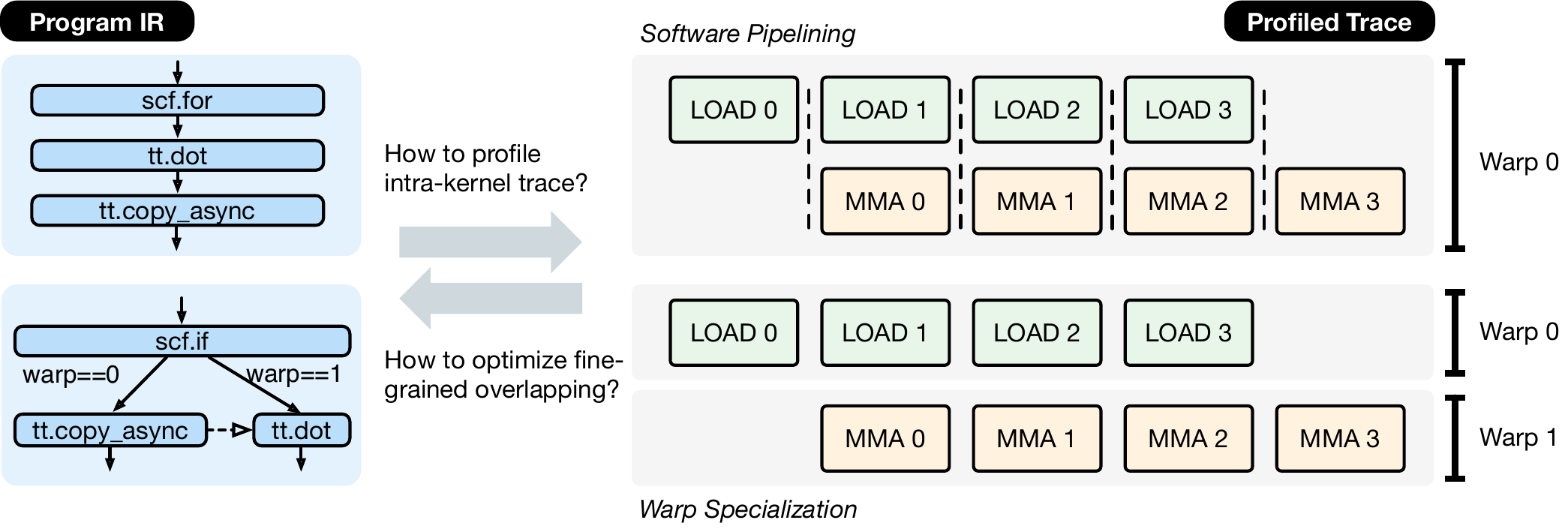}
    \vspace{-0.8cm}
    \caption{Motivating examples }
    \vspace{-1em}
    \label{fig:motivating_example}
\end{figure}

In this section, we analyze examples from current operator and compiler development that highlight the need for novel performance tooling, as illustrated in \Fig{\ref{fig:motivating_example}}.

\paragraph{Fine-grained Comprehension.} 
With the increasing complexity of heterogeneous execution units and asynchronous dataflow programming paradigms, it has become challenging to intuitively understand the execution pipeline. 
Users may choose SWP or WS to orchestrate data loading and computation for the cases shown in the example. 
However, to maximize hardware utilization, they must carefully determine stage partitions and the number of stages in SWP or decide execution orders and synchronization barriers in WS.
Existing profilers provide aggregated results that lack the program semantics needed to correlate profiled metrics. 
To truly understand overlapping behavior, it is essential to parse the loop structure, analyze cross-warp dependencies, and track fine-grained metrics throughout the execution. 
Users struggle to identify and address inefficiencies effectively.

\textit{\underline{Takeaway 1:}}  Performance profiling tools require the compiler’s IR to provide fine-grained performance metrics.

\paragraph{Performance Optimization.} 
Gaining a comprehensive understanding of execution behavior enables opportunities for manual optimizations. 
However, the challenge becomes even greater when implementing auto-optimizations within compiler passes. 
The compiler requires customizable performance feedback tied to the existing IR design to determine the optimal transformations.
As shown in the example, selecting the appropriate overlapping design of WS and SWP requires stable performance estimations for each method. 
The compiler pass must rely on precise profiling data to guide its optimization choices effectively. 
A programmable performance tool embedded within the compiler infrastructure provides the essential first step in enabling this workflow.

\textit{\underline{Takeaway 2:}} Compiler optimization passes need programmable performance profiling tools to effectively guide their optimization decisions.

%% file: section/3_compiler_centri_design.tex
\section{Compiler-centric Performance Tool} \label{sec:compiler_centric}

While it benefits compiler and kernel developers, the current compiler design lacks support for performance tooling. Existing MLIR dialects focus primarily on static, compile-time transformations and optimization passes, with little consideration for capturing dynamic traces or profiling semantics. This gap leaves developers without a native framework to analyze runtime behavior or fine-tune kernel performance within the compiler ecosystem.

\OURR{Designing performance tooling infrastructure as MLIR dialects is a non-trivial task. For example, in Triton, such dialects must bridge the abstraction gap between the high-level tensor IR and the low-level target-specific IR while remaining clean abstraction to be consistent and compatible with existing multi-level program IR semantics. The ideal infrastructure should be able to handle diverse profiling demands with great extensibility for ever-evolving GPU architectures.}
Addressing this complexity, we introduce the \thiswork{} dialect, the first of its kind, to integrate instrumentation passes directly into the compiler IR as a performance tooling infrastructure. 
The system is implemented upon the mainstream AI compiler Triton\cite{tillet2019triton}, which adopts MLIR as its compiler infrastructure.
We demonstrate the multi-level IR design and runtime convention of the \thiswork{} infrastructure as shown in \Fig{\ref{fig:implementation}}.
We further discuss several representative use cases with \thiswork{}.

\subsection{IR Design}

\begin{figure}
    \centering
    \includegraphics[width=\linewidth]{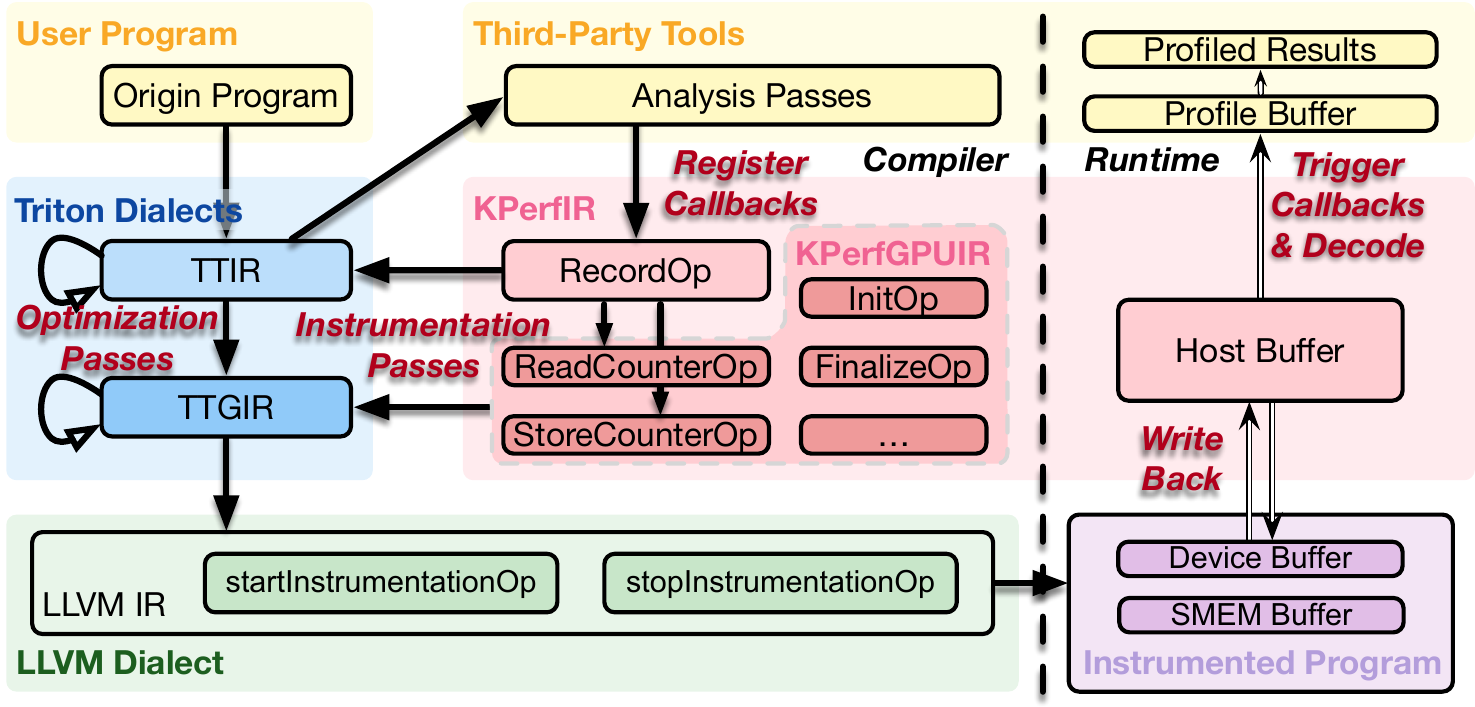}
    \vspace{-0.8cm}
    \caption{\R{IR design and conversion passes}}
    \vspace{-0.5cm}
    \label{fig:implementation}
\end{figure}

\R{
Within Triton's multiple-level IR structure, TTIR and TTGIR are the major program representations that developers interact with.
We implement the \thiswork{} as multi-layered MLIR dialects interacting with Triton's IR structure and link high-level MLIR operations to LLVM-IR level analysis.
As shown in \Fig{\ref{fig:implementation}}, we insert the record operations in MLIR dialects and lower them to architecture-specific TritonGPU operations. 
We also lower and instrument some scaffolding functions, such as data copy from local buffer to host memory, at the LLVM IR level. 
This architecture links the programming language constructs to hardware-related features for analysis. 
Specifically, we define novel operations at different levels to instrument profiling semantics and collect measurement data.
The rewritten Triton program is then lowered into LLVM IRs for further analysis and code generation.
This design balances programmability and complexity by higher-level IR design and lower-level code generation.
}

Instrumentation can be done at either the MLIR or LLVM IR level as a trade-off between flexibility and generality. 
MLIR level instrumentation can access high-level constructs, such as loops and data objects (e.g., tensors, matrices, etc.). 
However, the higher in the compiler pass pipeline instrumentation instructions are inserted, the more  accuracy and reliability are influenced by the compiler backend's instruction scheduling and reordering. 
In contrast, LLVM IRs are closer to the low-level assembly code and provide more information about GPU devices. 
Nevertheless, instrumentation at the LLVM IR level may lose connection to high-level program semantics.

\OURR{
As such, we design the compiler-centric performance tool by linking the high-level MLIR dialects to LLVM IR level analysis as shown in \Tbl{\ref{tbl:ir_design}}.
We provide the profiling instrumentation interface at both the TTIR and TTGIR levels since they are general representations that hide low-level vendor-specific details and capture key optimizations such as software pipelining, code motion, loop fusion, and warp specialization. 
\thiswork{} is our highest abstraction level, which includes the key \code{RecordOp} operator, representing a general program marker, whose semantic interpretation solely depends on the KPerfIR to KPerfGPUIR lowering pass configurations.
\code{RecordOp}'s inputs include \code{name} and \code{isStart}, specifying the annotation location and identifier in the program. An example of using \code{RecordOp} in the program is as follows.
}

\begin{table}[]
    \caption{\OURR{Main operations of \thiswork{}}}
    \vspace{0.5em}
    \label{tbl:ir_design}
    \resizebox{\linewidth}{!}{
    \begin{tabular}{llll}
    \Xhline{3.5\arrayrulewidth}
    Operation              & Inputs/Outputs                                                              & Attributes & Explanation                                      \\ \hline
    \rowcolor[HTML]{EFEFEF} 
    \multicolumn{4}{c}{\textbf{KPerfIR}} \\ 
    RecordOp               & \textbf{In:} name, isStart                                                       & -          & Main profiling IR.       \\ \hline
    \rowcolor[HTML]{EFEFEF} 
    \multicolumn{4}{c}{\textbf{KPerfGPUIR}} \\ 
    InitOp                 & \textbf{Out:} index\_ptr                                                   & \makecell[l]{BufferType,\\BufferStrategy}          & Initialize and allocate memory.                  \\
    \rowcolor[HTML]{EFEFEF} 
    FinalizeOp             &  \begin{tabular}[c]{@{}l@{}}\textbf{In:} index\_ptr,\\ \hspace{0.5cm}global\_ptr, data\end{tabular} & -          & \makecell[l]{Write back profiling to global memory\\ and add metadata.}       \\
    ReadCounterOp          &   \makecell[l]{\textbf{Out:} counter\_ptr}    & \makecell[l]{MetricType,\\Granularity}        & \makecell[l]{Read a GPU metric counter\\into a scalar register.}                    \\ 
    \rowcolor[HTML]{EFEFEF} 
    StoreCounterOp         &   \makecell[l]{\textbf{In:} counter\_ptr, index\_ptr}                      & isStart          & \makecell[l]{Store a metric counter into a buffer.}                    \\ \hline
    \multicolumn{4}{c}{\textbf{LLVM}} \\
    \rowcolor[HTML]{EFEFEF} 
    startInstrumentationOp & \begin{tabular}[c]{@{}l@{}}\textbf{In:} instrumentation pass,\\ \hspace{0.5cm}index\_ptr, buffer\_size\end{tabular} & -          & Trigger low-level instrumentation. \\
    stopInstrumentationOp  &  \textbf{In:} instrumentation pass                                         & -          & Stop low-level instrumentation.                  \\\Xhline{3.5\arrayrulewidth}
    \end{tabular}
    }
\end{table}

\vspace{-0.3cm}
\begin{figure}[h!]
    \centering
    \includegraphics[width=0.6\linewidth]{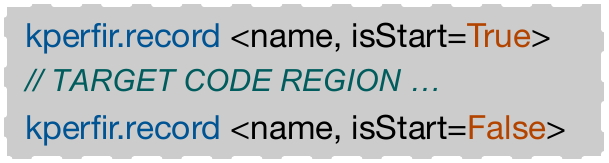}
    \vspace{-0.3cm}
    \caption{\OURR{An example of high-level record operations} }
    \label{fig:ir_example}
\end{figure}
\vspace{-0.3cm}

\OURR{
The \code{RecordOp} at KPerfIR level abstracts away the hardware detail and is lowered to GPU-specific KPerfGPUIR operations.
On the KPerfGPUIR level, we get vendor-independent but GPU-specific hardware features (e.g., shared memory abstraction).
When lowering to KPerfGPUIR, various MLIR pass options are given to determine the conversion, such as the \code{MetricType} (specifying the performance counter) and the \code{Granularity} (e.g., warp-group, warp, thread, etc). 
For example, when profiling GPU cycles, each \code{RecordOp} is lowered to a \code{ReadCounterOp} and a \code{StoreCounterOp}, which can be scheduled separately by the compiler for the proper program location. \code{ReadCounterOp} collects the exact GPU cycle register values. \code{StoreCounterOp} stores the counter value into the profiling buffer with an offset. 
The corresponding buffer and bookkeeping resources are determined and allocated during the lowering pass given configurable constraints.
}

\OURR{\Tbl{\ref{tbl:ir_design}} outlines key operators with inputs, outputs, and attributes presented in KPerfIR and KPerfGPUIR.
The concrete operations instrumented are controlled by various MLIR pass options in the lowering/conversion passes, including \code{BufferType}, \code{BufferStrategy}, \code{MetricType}, \code{Granularity} and resource constraints (e.g., buffer size). 
Given these controlling knobs, the compiler generates specific KPerfGPUIR operations for performance tools whose profiling locations are specified by the \code{RecordOp} markers.
For instance, specifying the \code{BufferStrategy} as circular, \code{MetricType} as clock, and \code{Granularity} as warp-group results in an intra-kernel profiler using a circular buffer to store the clock cycles of each warp-group (see \Sec{\ref{sec:tool}} for a detailed discussion of the profiler). In this case, the \code{StoreCounterOp} operation is converted to a \code{CircularStoreOp} to handle circular buffer event recording.
}

\OURR{
Besides, the compiler also generates scaffolding operations in KPerfGPUIR to setup and clean-up the environment of the profiling phase.
\code{InitOp} initializes the profiling states (e.g., buffer index). We use stack allocation to enable the LLVM toolchain to apply register promotion for the buffer index, which sits in the critical path of clock measuring. This operation returns the address of the allocated index and the memory load/store operations will be optimized away by keeping the index value in the register with LLVM backend's optimization. \code{FinalizeOp} performs the clean-up and writes back the profiled records. Similarly, the exact buffer memory allocation operations (\code{LocalAllocOp}, \code{GlobalScratchAllocOp}, and \code{StackAllocOp}) are determined by the configurations during the lowering from KPerfIR to KPerfGPUIR as well\footnote{Please refer to the documentation for more details at \url{https://triton-lang.org/main/dialects/ProtonOps.html}.}.
}

Lastly, we have the LLVM level markers \code{startInstrumentationOp} and \code{stopInstrumentationOp} to control the low-level library-based instrumentation.
The \code{startInstrumentationOp} tells the LLVM level to insert starting memory analysis/timestamp (including warp ID, SM ID).
And the \code{stopInstrumentationOp} inserts ending memory analysis/timestamp and flushes data from the local buffer to the scratch buffer/host pinned memory. This allows instrumentation at lower IR levels to be linked to upper-level data objects and flow structures (e.g., inserting timestamps at the LLVM IR level around data objects and flow structures that only exist at higher IR levels). 

Wrapping all instrumentation at the MLIR level makes the derived tools inextensible to ML frameworks due to their divergence on the supported operation set.
In contrast, inserting instructions at the LLVM IR level allows for flexible and reusable tools. 
Analysis code can be written in any language, compiled to LLVM bytecode, and inserted into MLIR framework code. 
This feature is particularly attractive in MLIR frameworks that are not linked to Clang or its libraries but include C++ instrumentation functions.
In other words, analysis functions can be written in HIP/CUDA and used in MLIR frameworks without C++ code (e.g., Triton, PyTorch).

\subsection{Compiler Passes}

We then leverage compiler passes to instrument the \code{RecordOp} to the target program and lower it to the underlying IRs.
There are two main levels of compiler passes handling their insertion and conversion, including lowering KPerfIR into KPerfGPUIR and lowering KPerfGPUIR into LLVM.

\OURR{
The KPerfIR to KPerfGPUIR lowering pass converts the \code{RecordOp} to \code{ReadCounterOp} and \code{StoreCounterOp}, and inserts operations for resource allocation (e.g., \code{LocalAlloc}), and setup/clean-up operations (e.g., \code{InitOp} and \code{FinalizeOp}). At the entrance of the kernel function, we allocate the profiling buffer, global scratch memory, and bookkeeping resources. We defer the discussion of memory management in \Sec{\ref{sec:profiler:memory}} and focus on the lowering pass explanation here. The allocated profiling buffer is associated with each \code{ReadCounterOp} to collect hardware performance counters (e.g., GPU cycle). The gathered raw data is written back to the GPU global memory by \code{FinalizeOp} in a pre-defined memory layout, which is inserted at the end of the kernel function. \code{FinalizeOp} has inputs including the number of the profiling records, base address of the profiling buffer, and base address of the profiler’s global memory.
}

\OURR{
Triton lowers TTGIR into the LLVM backend to generate vendor-specific code (e.g., ptx and amdgcn) and perform low-level optimizations, such as redundant code elimination~\cite{10.1145/582153.582167} and register promotion~\cite{10.1145/258916.258943}. We develop a set of LLVM conversion patterns to handle the code generation of the instrumented profiling operations. Specifically, the \code{InitOp} is lowered to a stack allocation (\code{llvm.alloca}) of the buffer index. The \code{ReadCounterOp} is lowered to a read of a performance counter (e.g., \%clock) to the register, and the \code{StoreCounterOp} is lowered to a register value store with tag creation and buffer index management. For the \code{FinalizeOp}, we compute the global memory offset for the current thread block and assign the first thread as a worker to write the entire profiling buffer back to the global memory.
}

We provide principal support regarding backends to reduce the interference of the high-level instrumentation in optimizations such as instruction re-ordering.
For Nvidia GPUs, the impact is negligible since the hardware is responsible for instruction scheduling with the PTX mostly following the instrumented program with \thiswork{}'s operators associated.
Instruction will not be scheduled around critical instruction like WGMMA.
AMD GPUs expose instruction scheduling to software, even for instruction SMEM load and MFMA in amdgcm, making the instrumented operator a key factor.
As such, we provide three levels of configurations adjustable by users: 1. manual adjusting with \thiswork{} hints; 2. direct instrumentation on amdgcn, and 3. specifying instruction scheduling window with barrier mask explicitly.

\begin{figure*}[t]
    \centering
    \includegraphics[width=\linewidth]{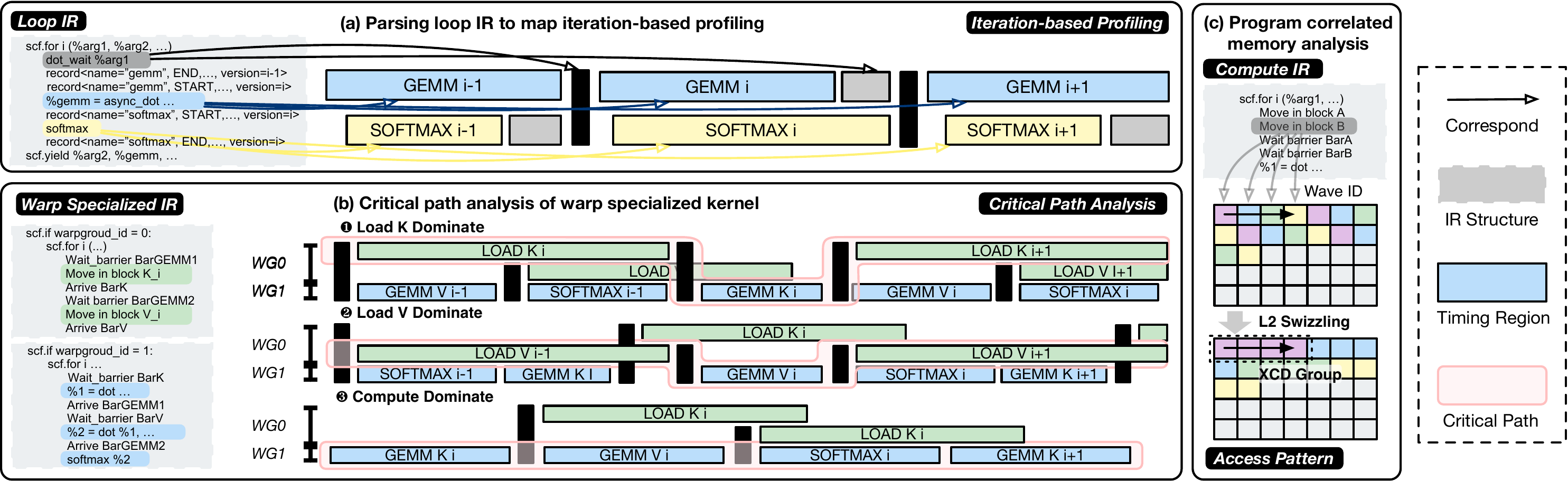}
    \vspace{-0.7cm}
    \caption{Novel use cases facilitated by \thiswork{}'s compiler-centric approach}
    \label{fig:compiler_centric_example}
\end{figure*}

\subsection{Working with Third-Party Tools} \label{sec:third_party}

We then demonstrate the interfaces for users to develop their third-party tools shown on the right side in \Fig{\ref{fig:implementation}}, including the instrumentation APIs handling rewriting and the necessary memory management to decode the profiled raw data. 

\paragraph{Instrumentation APIs.}
To build a customized performance tool with the \thiswork{} dialect, the user must explicitly modify the kernel signature and update the IR with instrumentation passes.
We provide two APIs for the user to interact with the \thiswork{}, the command-line API and the Python API.
With the command-line API, all Triton functions encountered at runtime with the command-line API are instrumented with the specified analysis passes.
This allows for handy manipulation of the target workload but lacks flexibility since users cannot skip kernels they are not interested in.

\R{
The Python API lets the user specify the instrumentation entry and the target function to be instrumented.
The instrumentation entry here refers to the insertion position in the compilation process (i.e., before/after which pass).
On the other hand, the instrumentation point specified by the \code{RecordOp} refers to the position in the IR.
It encompasses the core API as \code{KPerfIR.patch(instrumentation\_obj, fn)}, where \code{instrumentation\_obj} identifies the compiler pass to be instrumented (either before or after) and \code{fn} is optional to select the instrumented kernel.
If the \code{instrumentation\_obj} specifies MLIR instrumentation, we adopt MLIR's pass infrastructure \code{PassManagement}.
We will instrument LLVM at the end of the compilation passes.
Additionally, we provide a \code{KPerfIR.unpatch()} API to restore all existing instrumentation.
This involves maintaining the kernel's original and instrumented version within the \thiswork{} runtime.
}

\paragraph{Runtime Memory Management.}
Another important part of the third-party tool involves the management of buffer storage and profiling results.
At the top level, we rewrite the kernel signature with an extra last argument as a pointer to the device buffer and modify the calling convention.
The accumulated profiling data is then returned to the host buffer managed by the \thiswork{} runtime.
Here, we adopt a discarding-based circular buffer design, which will be elaborated with an actual tool design in \Sec{\ref{sec:profiler:memory}}.
We have two ways to manage GPU device buffer and host buffer communication, following prior studies~\cite{9355258,zhou2022valueexpert}
The first approach is to use CUDA-managed allocation so that the device buffer is allocated in a memory space visible to both the GPU and the CPU. 
The second approach is to allocate a priority stream with a higher priority than the one used by PyTorch's runtime and use that stream to copy the data back to the CPU.

Once the data is copied back to the CPU, it is decoded into a similar format to the CUPTI Activity API as a C/C++ struct. 
Third-party tools will register a callback to process this data when the runtime is initialized. 
Callbacks are triggered once the host buffer is decoded to allow these tools to process the data, like CUPTI activities. 
As such, the tools can post-process the data to compatible formats with their own front-end visualizers, such as Chrome Trace\cite{chrome_trace} or Hatchet\cite{bhatele2019hatchet}.

\subsection{Use Cases} \label{sec:usecase}

We then showcase several novel performance tools facilitated by \thiswork{} infrastructure as shown in \Fig{\ref{fig:compiler_centric_example}}.

\paragraph{Iteration-based Timing.} 
Understanding the synchronous behavior of warp groups across iteration boundaries is crucial for analyzing SWP. 
For instance, observing overlapping at specific iteration steps requires associating the loop induction variable \code{i} with a version argument in the timing records. However, traditional profiling tools lack access to the loop semantics necessary to determine iteration positions, making it challenging to comprehend the precise overlapping behavior of resources across iterations. 
By incorporating profiling passes into the compiler, we can instrument the IR to collect clock ticks alongside loop iteration index semantics, enabling post-processing to construct fine-grained timelines, as shown in \Fig{\ref{fig:compiler_centric_example}}-(a).
While tools like MLIR’s Python bindings can implement this functionality, they are ad-hoc and restricted to simple loop structures, underscoring the need for deeper integration of compiler semantics into profiling workflows.

\begin{figure*}
    \centering
    \includegraphics[width=\linewidth]{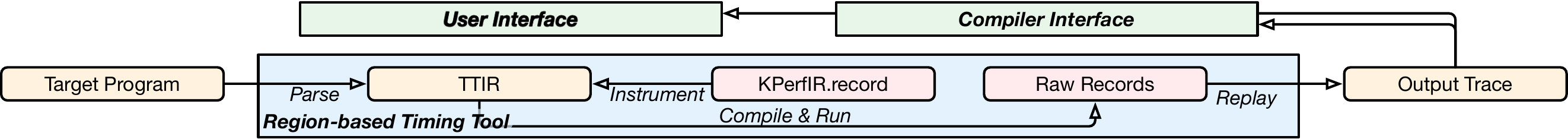}
    \vspace{-0.7cm}
    \caption{Workflow of the region-based timing tool}
    \vspace{-0.3cm}
    \label{fig:workflow}
\end{figure*}

\paragraph{Critical-path Analysis.}
Runtime feedback is invaluable for compiler passes, particularly for identifying execution critical paths in scenarios involving WS. 
For example, in the FA3 kernel, producer warp groups load input tensors, while consumer warp groups execute multiple GEMM operations and a softmax computation. 
The critical path—determined by factors like data movement and computation latency—varies with program tiling and access patterns, making static analysis insufficient for arrangement.
For the FA3 case, there are three common critical paths dominated by different operations, as shown in \Fig{\ref{fig:compiler_centric_example}}-(b).
By embedding profiling passes into the compiler, we can instrument timing records to capture runtime stage latencies and dynamically identify the critical path. 
This enables the compiler to strategically insert asynchronous barriers, reducing waiting bubble times and improving resource utilization. 
Such optimization passes highlight the value of integrating performance tooling with compiler passes.

\paragraph{Program Correlated Memory Analysis.}
Fine-grained memory profiling benefits significantly from a compiler-centric design, allowing developers to correlate memory performance metrics, such as access patterns, heat maps, and bank conflicts, with high-level program objects. 
Taking \Fig{\ref{fig:compiler_centric_example}}-(c) as an example, analyzing the memory access pattern of a specific data object, such as a tensor, requires parsing the program IR to insert profiling intrinsics that map high-level constructs to low-level hardware registers. This approach is particularly critical for optimizing L2-level scheduling, such as with AMD’s chiplet designs\cite{cnda_whitepaper}. 
Profiling warp-level memory access patterns and correlating them with program objects enables the design of optimized swizzling strategies for L2 access. 
Such capabilities, achievable only through compiler-centric profiling, demonstrate the necessity of integrating memory profiling tools into the compiler infrastructure.

%% file: section/4_profiler_tool.tex
\section{The Region-based Timing Tool} \label{sec:tool}
In this section, we dive into a powerful third-party performance tool built upon the \thiswork{} infrastructure that conducts region-based intra-kernel timing to assist in comprehending and optimizing the overlapping GPU hardware resources. 
To build a customized tool with \thiswork{}, the users must implement the instrumentation passes, the profile data management, and the post-processing process as introduced in \Sec{\ref{sec:third_party}}.
These are elaborated with the region-based timing tools as its workflow for instrumentation (\Sec{\ref{sec:profiler:workflow}}), its memory system (\Sec{\ref{sec:profiler:memory}}), and a trace replay post-processing (\Sec{\ref{sec:replay}}).

\subsection{Workflow} \label{sec:profiler:workflow}

We first introduce the workflow of the timing tool with two kinds of interfaces, the user interface and the compiler interface, as shown in \Fig{\ref{fig:workflow}}.
Depending on how the target program is parsed and the profiling regions are specified, different interfaces are invoked to instrument the profiling records.
\thiswork{} will then compile and execute the instrumented program to collect the raw performance counters.
At last, we replay the trace with the raw metadata and produce the final timeline trace.
This trace is output and feedback to different interfaces for visualization or profiler-guided compiler passes.

\paragraph{User interface} is the basic usage for developers to profile interested regions of a target GPU program.
We provide PythonDSL bindings in Triton kernels and also allow manual rewriting and overriding the dumped middle-level IR for the concerned regions.
It is straightforward as the developer manually rewrites the dumped middle-level IR for the concerned regions.
This is extremely useful when developing high-performance kernels and debugging efficiency problems.
The \thiswork{} will instrument the original program with rewritten program IR and produce the output trace.
Developers can utilize visualization tools, such as Chrome Tracer\cite{chrome_trace}, to get an intuitive view of the program timeline.

\paragraph{Compiler interface} is how the compiler interacts with the profiler.
Compared to the user interface, the compiler passes are responsible for parsing the middle-level IR and insert profiling record at concerned regions.
Similarly, the instrumented program is compiled and executed, and the profiled traces are feedback to the compiler passes within the compiler.

\subsection{Profile Data Management} \label{sec:profiler:memory}
\begin{figure}
    \centering
    \includegraphics[width=0.75\linewidth]{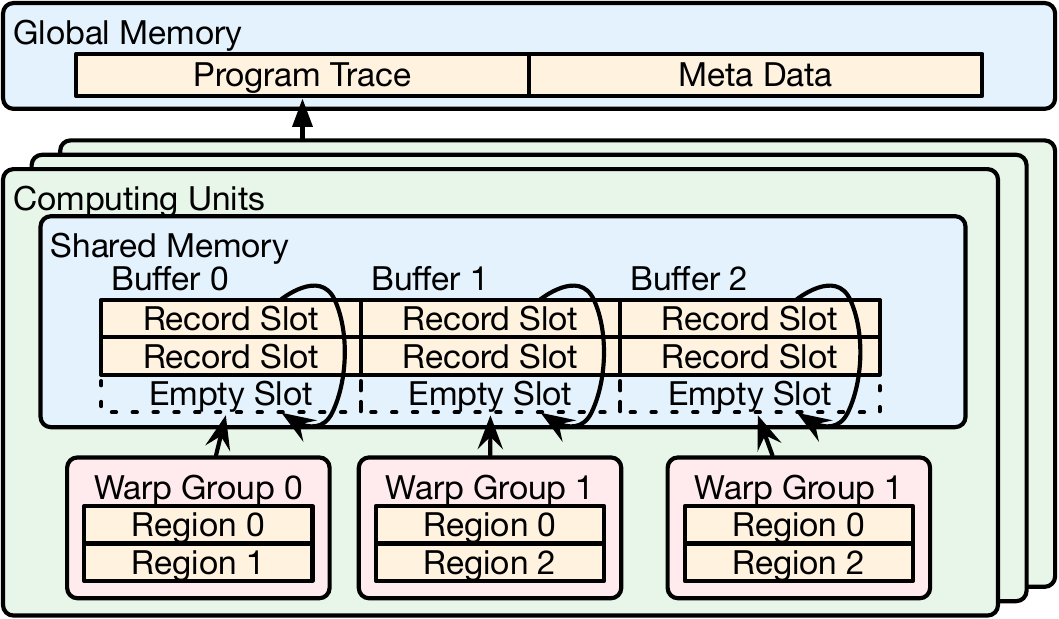}
    \vspace{-0.4cm}
    \caption{Memory management of the region-based tool}
    \label{fig:design}
\end{figure}

We then introduce the specific management and decoding techniques of the tool.
The system accommodating the storage at each memory hierarchy is shown in \Fig{\ref{fig:design}}.
We demonstrate the data structure and layout bottom-up in the following.

\begin{figure}
    \centering
    \includegraphics[width=\linewidth]{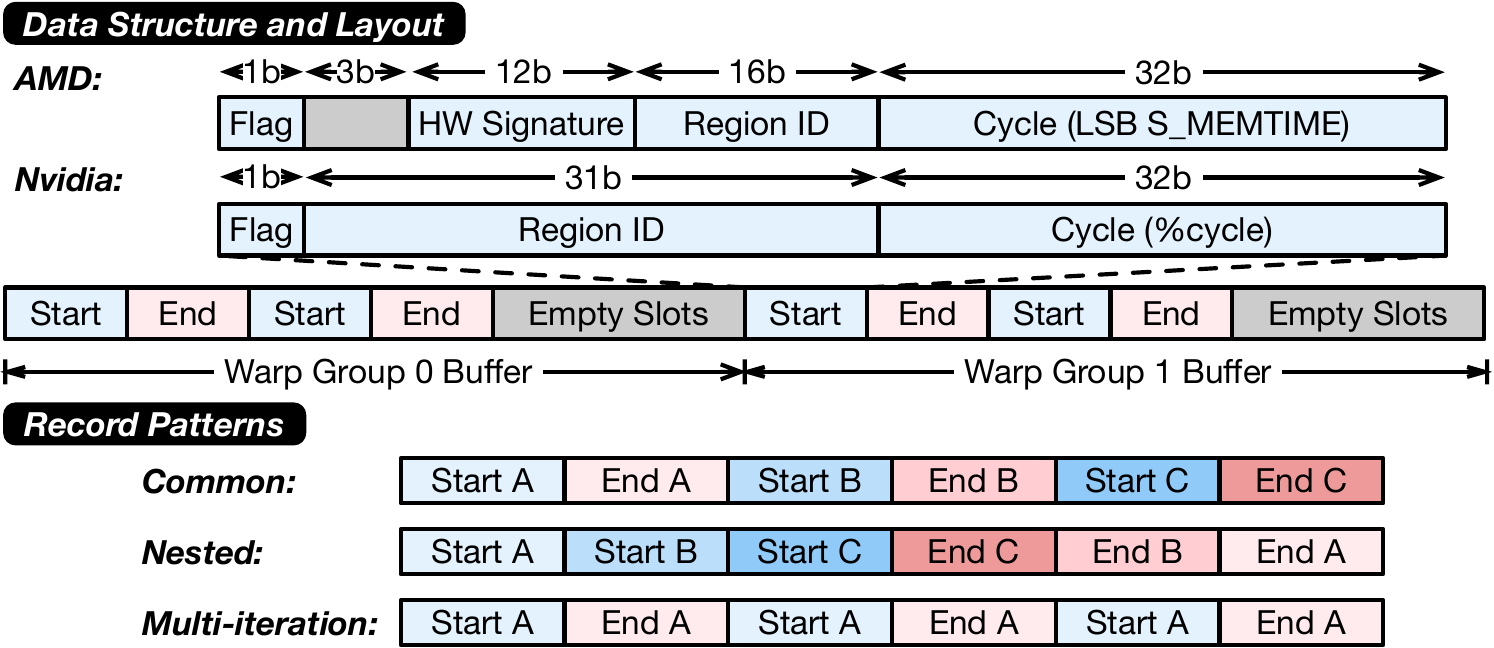}
    \vspace{-0.6cm}
    \caption{Circular buffer and record buffer}
    \label{fig:buffer}
\end{figure}

\paragraph{Warp-level Region Data Structure.}
For each fine-grained profiling region, there are two profiling records inserted, the region start record and region end record.
Each record has 8 bytes, containing a 4-byte tag and a 4-byte payload, as shown in \Fig{\ref{fig:buffer}}.
The tag contains 1 flag bit as \emph{START} or \emph{END}, and 31-bit control bits that are dependent on the backend will be explained later. 
The payload contains a 32-bit time stamp metric captured from the hardware counter (cycle for \%clock in Nvidia and the LSB 32 bits of \code{S\_MEMTIME} instruction in AMD). 
This design achieves a good trade-off between space capacity and write speed. 
Because the storage consumption is low and the store takes a vectorized store instruction.

\R{
We use a 32-bit clock to capture the cycle tick of the current record, which may cause value overflow.
We address this in the post-processing procedure, where we detect and throw exceptions on overflows.
The trace replay can compensate for the clock wrap-around overflow as long as each iteration runs less than 4 billion cycles (4 seconds under 1GHz). 
Since most of the execution is spent on loops, the restriction is relaxed to the loop iteration level, which is less than 1 millisecond for most cases, making 32-bit cycle data a safe choice.
If necessary, \thiswork{} could easily extend to a 64-bit clock by adding dedicated operators and attributes in the IR.
}

\paragraph{SM-level Data Layout.}
The record is then copied to the data buffer in shared memory from the local register file with a store instruction.
The data buffer is split into profiling spaces by warp groups with non-overlapping record slots, as shown in \Fig{\ref{fig:design}}. 
As such, the record index at compile time determines the store location offset in shared memory.
For example, if we allocate 64 slots (0.5KB) and each thread block contains 2 warp groups (8 warps, 256 threads), then each warp group has 32 slots. 
During compile time, we pre-compute the base address of each profiling space and generate code to manage the indexes for each warp group.

The memory layout of each warp group buffer is shown in \Fig{\ref{fig:buffer}}. 
Start and end records are interleaved in shared memory to support nested regions and multiple iterations. 
The lower part of the figure illustrates three profiling patterns: common, nested, and multi-iteration. 
Records are stored without pairing and aligned during trace replay to ensure correctness based on the profiling patterns.

\paragraph{Shared Memory Circular Buffer.}
Because shared memory has limited capacity, it may not be able to accommodate all profile data, especially with a long multi-iteration record pattern.
For example, with 2 warp groups, 4 profiled regions, and a loop with 512 iterations, it takes up to  4096 record slots, translating to a 16KB shared memory storage. 
For the production-level kernels, we usually get the available shared memory capacity left ranging from 1KB to 4KB for intra-kernel profiling's data buffer, only up to 1.75\% of a total of 228KB on H100 device\cite{choquette2023nvidia}. 

We propose a \emph{circular buffer} structure for each warp group's data buffer as a solution balancing accuracy and usability.
The circular buffer keeps only the trace's tail record cyclically when reaching the shared memory limit.
We overwrite the oldest results cyclically instead of flushing them to the lower-level memory.
The insight is that visualizing a few recent iterations is sufficient to identify the bottleneck.
We wrap around the index with a pre-calculated buffer capacity to implement the circular buffer.
If the captured events are overflowing the buffer, it is redirected to an occupied slot. 
This causes extra instructions for index management, which is addressed during compile-time.
We only need lightweight modular instructions to round the index at runtime.

\begin{figure}
    \centering
    \includegraphics[width=\linewidth]{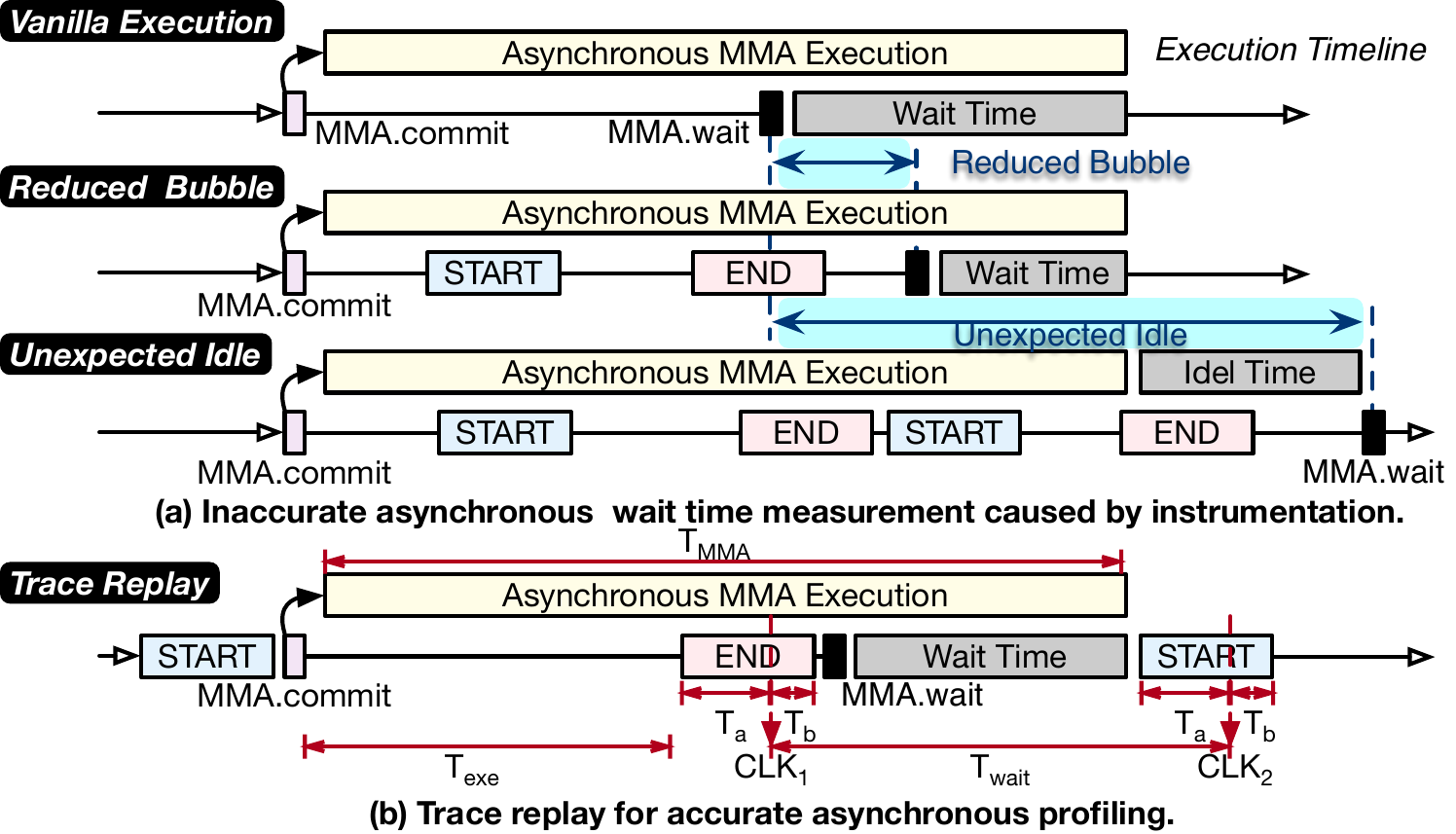}
    \vspace{-0.8cm}
    \caption{Trace replay}
    \vspace{-0.5cm}
    \label{fig:trace_replay}
\end{figure}

\R{
Besides, this \emph{circular buffer} is the default memory management behavior in the runtime.
The user can switch to the naive flush strategy, where the captured events are instantly written back to the global memory when the buffer is fully occupied.
This strategy can keep all the profile events with the overhead of many more frequent memory write instructions. 
In the \thiswork{}'s infrastructure, we provide a comprehensive operation to support the memory allocation at each hierarchy, including \code{Stack}, \code{Shared}, \code{Global}, and management strategies, including \code{Circular} and \code{Flush}.
Users can make use of these abstractions to develop their tool, combining the use of the buffers and management strategies.
}

\begin{figure*}
    \centering
    \includegraphics[width=\linewidth]{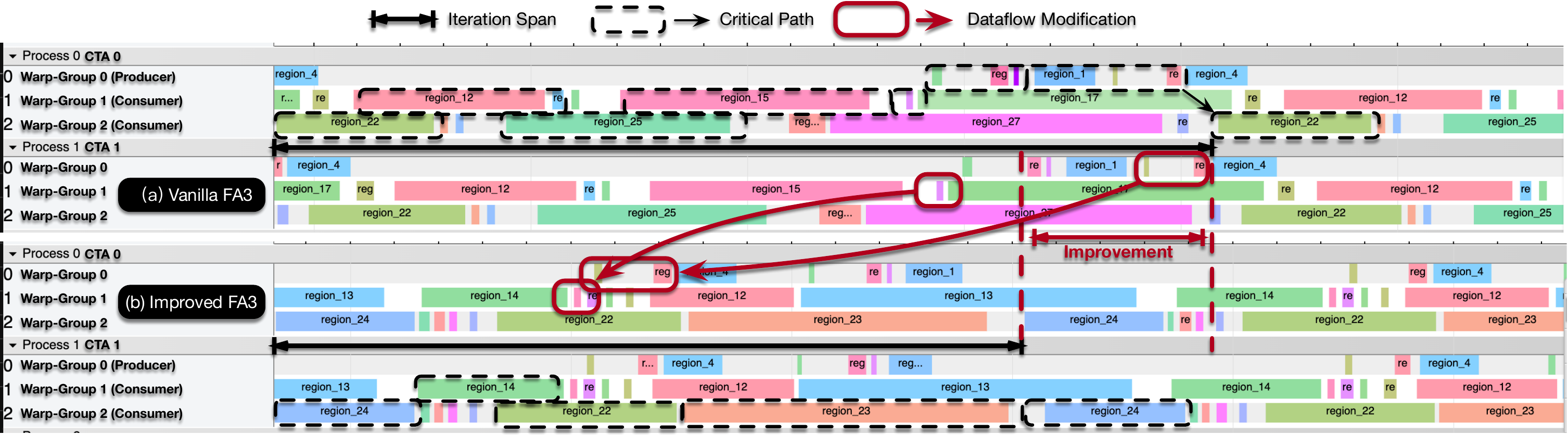}
    \vspace{-0.8cm}
    \caption{\OURR{Region-based timing results for FA3 kernels and overlapping improvements guided by profiling}}
    \vspace{-0.3cm}
    \label{fig:fa3_trace}
\end{figure*}

Specifically, we implement a collaborative store strategy for the AMD backend to minimize profiling overhead. 
Since only one timestamp record is profiled per warp or warp group, AMD GPUs utilize branching with thread masks to issue instructions with predication. 
However, this thread divergence can lead to unexpected instruction cache misses, causing overheads up to 600 cycles. 
To mitigate this, we enforce all threads to write to the same shared memory location, retaining only the last result. 
To align the records, we include an additional 12-bit signature derived from the least significant bits (LSB) of the \code{HARDWARE\_ID REG}, as shown in \Fig{\ref{fig:buffer}}. 
This signature encodes the \code{wave\_slot\_id}, \code{SIMD\_id}, and \code{pipe\_id} to annotate the profiled thread. 
At the warp group level, we ensure all four warps write to the same shared memory location.

\paragraph{Global Program Trace.}
Lastly, we gather the profiling data from all SMs back to the GPU global memory before returning from the kernel function. 
We keep the temporary profile data in the shared memory and copy the entire data buffer at once by the end of the kernel. 
Several extra metadata are attached for reconstructing the trace as shown in \Fig{\ref{fig:design}}, including thread block index, warp-level indices, and a number of recorded slots in each warp group. 
The Triton frontend automatically patches the kernel function with an extra argument (\code{profile\_mem}), allocating the GPU scratch memory for such profile data.

\subsection{Trace Replay} \label{sec:replay}

The major challenge in the region-based timing scenario of intra-kernel profiling is the perturbations caused by the profiling instructions.
Post-processing techniques, such as periodic sampling\cite{sample_based_profiling}, are usually used to mitigate the inaccuracies by sampling the performance metrics periodically and summarizing the results afterward to reduce the overhead associated with data collection.
\R{
In the region-based timing tool, we propose a similar post-processing method \emph{trace replay} to rescale the profiled regions according to the program semantics owing to the \thiswork{}'s design.
We can get accurate visualized results using this trace replay mechanism by offsetting a constant to account for the clock measuring overhead.
}

For regions with only synchronous instructions, we can easily rebuild the region by subtracting the overhead of the record operation.
For asynchronous instructions, we have two kinds of common inaccuracies as shown in \Fig{\ref{fig:trace_replay}}.
In the vanilla execution timeline example, we have a sequence of successive instructions executed after issuing an asynchronous MMA operation on the Tensor Core.
After the program reaches the barrier in the instruction stream, there is a period of waiting time before the functional unit can produce results.
To understand the asynchronous behavior here, users can insert profiling records in the successive execution to better align the operations.
However, the introduction of such records causes inaccurate wait time measurements.
In the \emph{reduced bubble} case, the wait time is underestimated due to the extra overhead caused by profiling.
Things are even worse in the \emph{unexpected idle} case, where the functional unit's execution time cannot cover the profiling overhead.
An unexpected idle time will be observed, and the optimization will be misleading.

\begin{table}[]
  \caption{Profiled regions of the FA3 kernel}
  \label{tbl:fa3}
  \centering 
  \resizebox{\linewidth}{!}{
  \begin{tabular}{llll}
  \Xhline{3\arrayrulewidth}
  \multirow{2}{*}{Tag} & \multirow{2}{*}{Function}     & \multicolumn{2}{c}{Region ID} \\ \cline{3-4} 
                       &                               & Vanilla       & Improved      \\ \hline
  \rowcolor[HTML]{EFEFEF} 
  Load K               & Load the K tensor.            & 3      & 3       \\
  Load V               & Load the V tensor.            & 6      & 6       \\
  \rowcolor[HTML]{EFEFEF} 
  GEMM0                & Compute QK on TC.             & 12,22         & 12,22         \\
  GEMM1                & Compute PV on TC.             & 14,24         & 15,25         \\
  \rowcolor[HTML]{EFEFEF} 
  Softmax              & Compute softmax on CUDA core. & 13,23         & 17,27         \\ \Xhline{3\arrayrulewidth}
  \end{tabular}}
  \vspace{-0.5cm}
  \end{table}

To address this, we propose deducting the accurate wait time after profiling in the trace replay procedure as shown in \Fig{\ref{fig:trace_replay}}-(b).
Instead of placing one \code{END} record after the asynchronous instructions, we insert two \code{START} records before the asynchronous launch and after the wait barrier and one \code{END} record right before the barrier.
This guarantees an accurate wait time $T_{wait}$ measurement as $T_{wait} = (CLK_2-T_a) - (CLK_1-T_a)=CLK_2-CLK_1$, where $T_a$ and $T_b$ are the elapsed time in the record region before and after the clock read instruction respectively.
The profiling overhead is canceled out in the post-processing with the two carefully placed records.
We can derive the $T_{exe}$ time similarly.
This method requires the $T_{MMA}-T_{exe}>T_a+T_b$.
Since the profiling overhead is less than 25 cycles for most cases, as shown in \Sec{\ref{sec:evaluation:cycle}}, the execution time of the functional unit can cover only the record inserted here, which is usually about 1000 cycles.

%% file: section/5_evaluation.tex
  \begin{figure*}
    \centering
    \includegraphics[width=0.9\linewidth]{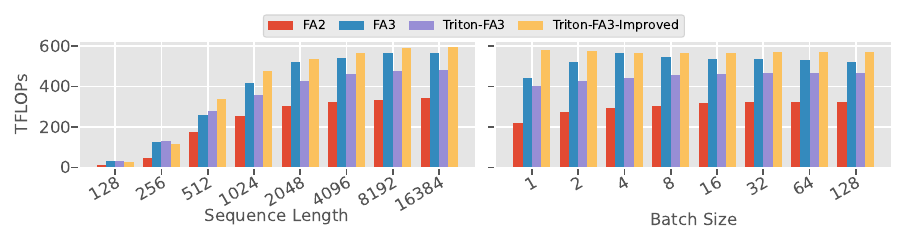}
    \vspace{-0.6cm}
    \caption{Benchmarking FA3 kernels with a head dimension of 128 and 16 heads. The batch size and sequence length are set to 16 and 4096 in the two benchmarks, respectively.}
    \label{fig:fa3_benchmark}
  \end{figure*}

\vspace{-0.3cm}
\section{Evaluation} \label{sec:evaluation}
\vspace{-0.3cm}

In this section, we evaluate the performance of the designed region-based timing tool and the \thiswork{} approach.

\vspace{-0.3cm}

\subsection{Experimental Setup}

\paragraph{Testbed.} 
The evaluation of \thiswork{} was conducted on servers equipped with state-of-the-art GPU cards, including NVIDIA H100-HBM3 and AMD MI300X. The software environment included Triton 3.0.0 and LLVM 19.1, ensuring compatibility with the latest compiler technologies and GPU architectures.

\paragraph{Benchmarks.}
To assess the usability and performance of the proposed tool, we benchmarked it on key AI workloads, specifically GEMM\cite{Triton-gemm} and experimental Flash Attention operators from Triton\cite{Triton-fa3}. 
These operators are foundational to popular AI models, representing the majority of workloads in tasks such as training and inference. 
We selected mainstream implementations of these operators that deliver state-of-the-art performance and conducted sensitivity experiments to evaluate the method's effectiveness and scalability.

\subsection{A Journey of Flash-Attention 3} \label{sec:use_case}

We first demonstrate utilizing the region-based timing tool to improve the FA3 kernel.
As FA3 involves many optimization techniques dedicated to the Nvidia platform, we focus on optimizing the H100 GPU first and conduct many more thorough benchmarks later.

\subsubsection{Attention Overlapping Optimization}

We begin by demonstrating the usage of the region-based timing tool to identify a particular inefficiency in the FA3 kernel's WS overlapping design. 
Using the proposed timing tool, we parse the FA3 kernel’s IR structure and profile its asynchronous behavior to guide a better overlapping design. 
The critical stages of the FA3 kernel are identified and listed in \Tbl{\ref{tbl:fa3}}. Each stage is labeled with a unique region ID, providing a clear mapping for analysis.
By employing the timing tool, we obtain a fine-grained timeline trace of the vanilla FA3 kernel from Triton, shown in \Fig{\ref{fig:fa3_trace}}-(a), which also illustrates the user interface with the Chrome Trace as the front-end.
The trace identifies the critical path consisting of 4 GEMMs and 2 loading stages, including region 22, 12, 25, 15, 6 and 3.
Specifically, the loading V stage of region 6 is blocked by the arrival barrier of region 16 in consumer 1, causing a longer critical path and lower hardware utilization.

Following this observation, we can advance the arrival barrier of V in region 16, as shown with the red arrows, to prevent it from stalling the issue of successive data loading.
The GEMM1 regions are released from the critical path by fully overlapping it with the concurrent K tensor loading.
This modification breaks data dependency between the arrival and computation and requires extra pre-loading in the prologue before the iterations.
We can achieve a much more compact timeline where the softmax and GEMM computation are overlapped, shown in \Fig{\ref{fig:fa3_trace}}-(b).
The improved overlapping results in a reduced wall time of each iteration in the kernel.

  \begin{table}[]
    \caption{Performance models}
    \vspace{0.5em}
    \label{tbl:performance_model}
    \resizebox{\linewidth}{!}{
    \begin{tabular}{lm{7cm}}
    \Xhline{3\arrayrulewidth}
    Category      & Analytic Model \\ \hline
    SWP Model     & \vspace{-0.5cm}\[
    \begin{aligned}
    \Delta = &N_{WG}*N_{pipe}*\Sigma_i T_{comp}-Max_i(T_{load}^i+T_{comp}^i) \\
    &\begin{cases}
    \Sigma_i T_{comp}^i * N_{loop}, \Delta >= 0 \\ 
    Max_i(T_{load}^i+T_{comp}^i)\times N_{loop} / N_{pipe}
    \end{cases}
    \end{aligned}
    \]      \vspace{-0.3cm}         \\
    WS Model      &   $\Sigma_{i \in CriticalPath} T_{load/comp}^i$             \\
    Compute Model &  $ FLOPs / Throughput $              \\
    Memory Model  &   $T_{read} + Bytes / Bandwidth$              \\ \Xhline{3\arrayrulewidth}
    \end{tabular}
    }
    \end{table}

\subsubsection{Performance Modeling for GPU Overlapping}

We then demonstrate how to adopt the proposed compiler-centric profiler design to build a compiler pass to determine the optimal overlapping design.
This is achieved by extracting the critical path of the FA3 kernel with the profiling results and compute the utilization of the workload using an overlapping performance model. 
The performance model is outlined in \Tbl{\ref{tbl:performance_model}}, focusing on compute and data loading stages while simplifying less relevant aspects such as initialization, epilogue, and overall kernel performance, which have been studied extensively in prior works~\cite{huang_alcop_2023}.We emphasize the non-trivial parts of modeling the overlapping techniques.

For the SWP model, the discriminant $\Delta$ determines whether the bottleneck lies in data loading or computation.
If $\Delta \geq 0$, the data loading latency is fully overlapped by computation, and the total latency corresponds to the accumulated computation time across all stages.
Conversely, if $\Delta < 0$, the latency is dominated by the most time-consuming loading and computation stages.
For the WS model, the latency is determined by identifying the kernel’s critical path, as discussed in \Sec{\ref{sec:usecase}}. The WS latency is intuitively calculated as the sum of the latencies of all stages along the critical path.

Using this model, we can quantitatively assess the overlapping efficiency of the FA3 kernel. The region-based profiler collects detailed performance data, allowing us to analyze the current implementation and refine overlapping designs to minimize latency and improve resource utilization.
For example, for an attention kernel with head dimension 128, head number 16, batch size 16, and sequence length 4096, the 2-stage SWP is calculated with a 467.07 TFLOPs and the vanilla Triton FA3 reaches a 526.97 TFLOPs.
With the optimization passes incorporated, we can tune an FA3 kernel with improved overlapping, achieving 582.44 predicted TFLOPs.

\subsubsection{Improved FA3 Evaluation}

Combining the manual optimization passes with the profiling insights and performance model pass together, we can get an improved FA3 compiler pass optimization suite built upon Triton.
Compared to the original execution dataflow of the vanilla Triton FA3, the optimized FA3 kernel achieves a $24.1\%$ improvement as shown in \Fig{\ref{fig:fa3_benchmark}}. 
This shows the efficacy of the optimization insights discovered with the assistance of the performance tool.
The improved Triton-FA3 kernel also outperforms the best manual FA3 kernel\cite{shahflashattention} by $7.6\%$. 
Particularly, without the assistance of the region-based timing tool, the vanilla Triton-FA3 kernel fails to achieve a competitive performance compared to the manual kernels. 
This suggests the immeasurable value of a fine-grained profiling tool benefiting the compiler passes and kernel optimization.

\begin{figure}
    \centering
    \includegraphics[width=0.5\linewidth]{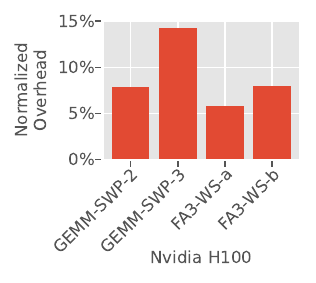}~
    \includegraphics[width=0.5\linewidth]{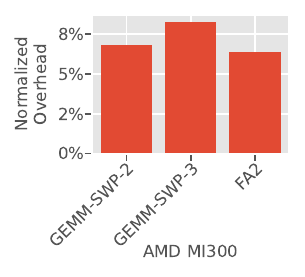}
    \vspace{-0.85cm}
    \caption{Normalized latency overhead}
    \label{fig:latency_overhead}
\end{figure}

\subsection{Profiling Overhead}
We evaluate the profiling overhead of the proposed region-based timing tool in terms of both latency and memory consumption. Particular attention is given to shared memory usage, which is a critical bottleneck in the design of performance tools for GPU workloads.
We benchmark SOTA GEMM operators with SWP designs divided into 2 or 3 stages, marked as GEMM-SWP-2/3, and WS FA3 operators with vanilla overlapping and the improved counterpart with the profiling insights, marked as FA3-WS-a/b.

\paragraph{Latency Overhead.}
We measure the end-to-end latency of the instrumented kernels and normalize the results to their original execution time. \Fig{\ref{fig:latency_overhead}} illustrates the latency overhead across the evaluated benchmarks. For most cases, the overhead remains under 10\%, ensuring that the proposed tool is practical for real-world scenarios without significantly impacting kernel performance. 
For the most complicated SWP GEMM kernel with three stages, we insert many records to cover its three stages.
Even in this case, the overhead is kept within 15\%.
This minimal latency impact is crucial for enabling real-time profiling and maintaining production-level throughput during optimization workflows.

\R{
We hightlight that with the post-processing trace replay technique in \Sec{\ref{sec:replay}} mitigating the interference of the profiling instructions, we can get relatively accurate profile results even if the overhead is extended. 
We have also introduced several memory management techniques like circular buffers, which together reduce the profiling overhead to approximately 8\% in our evaluation. Moreover, ongoing enhancements, such as low-level instruction scheduling and clock offsetting, can further reduced the overhead.
}

\paragraph{Shared Memory Overhead.}
As detailed in \Sec{\ref{sec:profiler:memory}}, the proposed timing tool employs a circular buffer design to store recent profiling records within the limited shared memory available on GPUs. This approach ensures compatibility with diverse workloads, regardless of the size of the available shared memory, while also allowing flexible adaptation of the profiling life cycle.

\begin{figure}
    \centering
    \vspace{-0.3em}
    \includegraphics[width=0.5\linewidth]{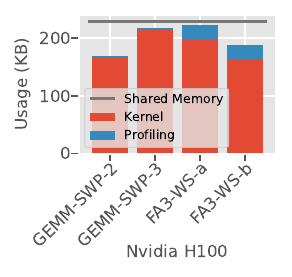}~
    \includegraphics[width=0.48\linewidth]{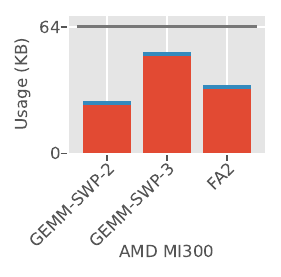}
    \vspace{-1cm}
    \caption{Memory usage}
    \label{fig:memory_overhead}
\end{figure}

\Fig{\ref{fig:memory_overhead}} shows the shared memory usage across the benchmarked workloads. For this experiment, the circular buffer size was set to either maximize usable shared memory space or accommodate all profiling records for the workload. The results demonstrate that the tool effectively operates within the constraints of shared memory, even for industrial-grade kernels, without spilling profiling records to external memory.

For instance, in the most storage-intensive kernel, the SWP GEMM with 3 stages, there remains an unused shared memory space of 10.9 KB. With 4 profiled regions, the timing tool can track up to 16 iterations, providing sufficient and stable coverage to estimate the entire execution process. This efficient memory usage highlights the applicability of the tool for large-scale GPU workloads.

\subsection{Low-level Deep-dive} \label{sec:evaluation:cycle}

To provide an empirical analysis of the optimization performance degradation introduced by \thiswork{}, we examine its impact at the low-level instruction level. Specifically, we identify the instrumented instructions generated by IR-level profiling and compare them with their vanilla counterparts to evaluate how profiling IR affects the compilation stack.

\begin{figure}
\centering
  \begin{minipage}[t]{0.37\linewidth}
    \centering
    \includegraphics[width=\linewidth]{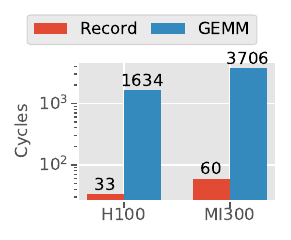}%
    \vspace{-0.5cm}
    \caption
      {%
Cycle-level benchmarks
      }
      \label{fig:cycle_benchmark}
  \end{minipage}
  \hfill
  \begin{minipage}[t]{.57\linewidth}
    \centering
    \vspace{-2cm}
\resizebox{\linewidth}{!}{
\begin{tabular}{lrrr}
\Xhline{3\arrayrulewidth}
                     & GEMM   & Theoretical & Actual                     \\ \hline
\makecell[l]{\small{Active}\\\small{     Cycles}}        & 199381 & 224981      & \multicolumn{1}{r}{229663} \\
\makecell[l]{\small{Relative}\\\small{     Performance}} & 0.89   & 1           & 1.02 \\   \Xhline{3\arrayrulewidth}                  
\end{tabular}
}
\vspace{-0.02cm}
    \captionof{table}
      {%
        Performance degradation evaluation
        \label{tbl:degradation}%
      }
  \end{minipage}
\end{figure}

\paragraph{Cycle-level Overhead.}
At the GPU assembly code level, the SASS ISA, each \thiswork{} record node is lowered to three instructions: a clock read instruction, an integer move instruction, and a predicated store instruction. These instructions collectively incur an average overhead of 33 cycles, as shown in \Fig{\ref{fig:cycle_benchmark}}. This latency represents the per-record profiling overhead observed in our sensitivity analysis. For loop-based timing, where profiling records are inserted into loop structures, five additional instructions are generated at the loop entry for index management. The instructions for each profiling record remain consistent at three, with their destination adjusted dynamically based on the loop iteration variable.

\paragraph{Optimization Degradation.}
One critical observation in our analysis is that profiling instrumentation can interfere with the compiler’s low-level optimizations, such as instruction reordering and constant folding. Integrated profiling semantics at the IR level inherently trade some low-level control for improved tool portability and compatibility, as discussed in \Sec{\ref{sec:compiler_centric}}. This tradeoff introduces potential risks of unintended instruction reordering when profiling instructions interact with adjacent non-profiling instructions.

To study this impact, we modeled the theoretical execution time of the instrumented kernel as
\begin{equation}
    T_{theoretical} = T_{vanilla} + N_{record} * Cycle_{record}.
\end{equation}
The $N_{record}$ is the number of instrumented records, and the cycles are measured as \Fig{\ref{fig:cycle_benchmark}}.
Using this model, we compared the theoretical execution time with the actual performance of the instrumented workload to assess deviations caused by profiling instrumentation.

The results, summarized in \Tbl{\ref{tbl:degradation}}, show that the performance impact remains within 2\%. This suggests that the profiling instrumentation at the compiler IR level introduces minimal overhead while maintaining compatibility with compiler optimizations. The findings highlight the practicality of the proposed approach, ensuring accurate profiling without significant performance degradation.

%% file: section/6_discussion.tex
\section{Discussions}

\subsection{Limitations}
\R{
One limitation of our approach stems from restricted visibility into certain vendor-specific performance counters. Since \thiswork{} inserts profiling instructions directly into the program to collect runtime information, the set of accessible metrics is constrained by the vendor’s ISA. In contrast, vendor-provided tools such as NVIDIA’s Nsight Compute (NCU) and AMD’s ROCm Tools (e.g., ATT) have privileged access to undocumented performance registers not exposed to third-party developers. These proprietary counters can only be utilized through their official APIs or libraries. While our infrastructure may not achieve full parity with these tools in terms of raw counter access, our IR-level instrumentation enables unique insights into program behavior otherwise difficult to obtain with closed-source profilers.
}

\subsection{Workload Generality}

\R{
\paragraph{Beyond AI Workloads.} While \thiswork{} currently targets AI workloads through its integration with the Triton compiler, the core ideas behind our approach, namely a multi-level IR and transformation passes for performance tools, are broadly applicable. These techniques can generalize to other domains and compilers, such as those used in high-performance computing (HPC)~\cite{reed2022reinventing} or scientific simulation~\cite{keyes2013multiphysics}.
}

\R{
\paragraph{Distributed Workloads.} \thiswork{} is also amenable to distributed GPU workloads~\cite{wang2025wlb,rap}, which are typically structured as a series of kernels involving both computation and communication. For workloads that launch computation and communication kernels separately, \thiswork{} already provides native support. 
For fused kernels, where computation and communication are fused within a single kernel\cite{chang2024flux,punniyamurthy2024optimizing}, \thiswork{} can theoretically instrument and analyze them as well. 
However, full integration depends on ongoing upstream efforts to extend Triton’s support for such fused distributed execution.
}

\section{Conclusion}

As AI compilers and GPU architectures continue to evolve, performance tooling must also advance to meet the growing demands of modern compilers. 
To address this need, we introduced \thiswork{}, a novel compiler-centric profiling infrastructure built upon the Triton compiler. 
\thiswork{} bridges the gap between compiler and profiler design, enabling the development of programmable, reusable tools and unlocking new possibilities for performance analysis and compiler optimizations. 
We envision \thiswork{} as a foundation for an open, compiler-centric ecosystem, empowering the community to construct diverse and innovative performance tools that adapt to the dynamic needs of AI workloads.

\section*{Acknowledgement}

The authors would like to thank the anonymous reviewers for their constructive feedback on improving the work.
We also thank our shepherd for the support during the revision process.
We would like to thank our colleagues from Meta: Bert Maher, Hongtao Yu, Taylor Robie, Elliot Gorokhovsky and OAI people Philippe Tillet, Thomas Raoux, Pawel Szczerbuk, for their early discussion and feedback on this project.
This work was inspired by Adam Paszke's work on MosaicGPU, presented both at Meta and Triton Conference.
This work was supported in part by NSF 2124039.
Keren Zhou's work is supported by a donation from AIGCSEMI LLC.


%% file: section/artifact.tex
\section{Artifact Appendix}

\subsection*{Abstract}

The \thiswork{}  is a performance tool infrastructure for the Triton~\cite{tillet2019triton} compiler and  the results for the OSDI'25 submission are derived from some feature branches.
While the core concept is stable, the implementation is still evolving and subject to change.
A formal documentation can be found at \url{https://triton-lang.org/main/dialects/ProtonOps.html}.

\subsection*{Scope}

We showcase the usability of the performance tool and the improvement results discussed in the paper by reproducing \Fig{\ref{fig:fa3_trace}} and \Fig{\ref{fig:fa3_benchmark}}.
Specifically, profiling the FA3 kernel with the region-based performance tool demonstrate the usage of the compiler-centric design.
A comparison of the profile trace from the vanilla and improved FA3 kernels shows the methodology of intra-kernel region profiling.
And the evaluation results shows the actual performance improvement.

\subsection*{Contents}

We provide a Docker image that contains the Triton compiler and evaluation scripts for the artifact evaluation.
Please follow the installation instructions in the following section to set up the environment.
Within the Docker image, the contents are organized as follows:

\vspace{1em}
workspace/

\quad{}----- triton               \hspace{5em} \textcolor{ForestGreen}{\% Triton compiler source code}

\quad{}----- tritonbench          \hspace{2.6em} \textcolor{ForestGreen}{\% Triton FA3 benchmark suite}

\quad{}----- kperfir\_artifact    \hspace{1.2em} \textcolor{ForestGreen}{\% Other scripts and files}

\subsection*{Hosting}

This artifact is open sourced at \url{https://github.com/ChandlerGuan/kperfir_artifact} with the main branch and commit version \texttt{e45891d}.

\subsection*{Requirements}

The Docker environment requires a Linux-based system with NVIDIA Container Toolkit installed.
While the \thiswork{} project is designed to be cross-platform, the artifact evaluation for the FA3 kernel requires a Nvidia H100 GPU.